\documentclass{emulateapj}
\pdfoutput=1
\shorttitle{Flip-Flop Instability}
\shortauthors{Blondin and Pope}

\begin{document}

\title{Revisiting the ``Flip-Flop" Instability of Hoyle-Lyttleton Accretion}

\author{John M. Blondin and T. Chris Pope}
\affil{Department of Physics, North Carolina State University, Raleigh, NC 27695-8202}
\email{John\_Blondin@ncsu.edu}


\begin{abstract}

We revisit the flip-flop instability of two-dimensional planar accretion using high-fidelity numerical simulations.  
By starting from an initially steady-state axisymmetric solution, we are able to follow the growth of this 
overstability from small amplitudes.  In the small-amplitude limit, before any transient accretion disk is 
formed, the oscillation period of the accretion shock is comparable to the Keplerian period 
at the Hoyle-Lyttleton accretion radius ($R_a$), independent of the size of the accreting object.  
The growth rate of the overstability increases dramatically with decreasing size of the accretor, 
but is relatively insensitive to the upstream Mach number of the flow.  We confirm that the flip-flop 
does not require any gradient in the upstream flow.  Indeed, a small density gradient 
as used in the discovery simulations has virtually no influence on the growth rate of the overstability.  
The ratio of specific heats does influence the overstability, with smaller
$\gamma$ leading to faster growth of the instability.  For a relatively large accretor (a radius of $0.037 \,R_a$)
planar accretion is unstable for $\gamma = 4/3$, but stable for $\gamma \ge 1.6$.  Planar accretion
is unstable even for $\gamma = 5/3$ provided the accretor has a radius of $< 0.0025\, R_a$. 
We also confirm that when the accretor is sufficiently small, the secular evolution is described
by sudden jumps between states with counter-rotating quasi-Keplerian accretion disks.

\end{abstract}

\section{Introduction}

The instability of hydrodynamic accretion from a moving medium and its relevance to the temporal behavior of accreting x-ray sources has been debated since the discovery of a flip-flop instability in two-dimensional hydrodynamic simulations of an accreting compact object \citep[hereafter FT]{ft88}.  
This non-axisymmetric instability has been subsequently seen in numerous hydrodynamic simulations of two-dimensional planar accretion
\citep{mss91,zwn95,blt97,shima98,pom00}, but has not been unambiguously identified in three-dimensional simulations \citep{r99}.
Given that accretion-powered X-ray sources in Be-type binary systems are expected to accrete from a planar disk expelled by the companion Be star, attention has focussed on the relationship of the flip-flop instability to these Be-type X-ray pulsars.  
\citet{tfb88} interpreted the recurrent flares in EXO 2030+375 in terms of the episodic accretion resulting from the flip-flop instability.  \citet{blt97} argued that the variable accretion torques associated with the flip-flop instability may explain the erratic spin behavior in these systems.  

The first quantitative description of hydrodynamic accretion was given by \cite{hl39a}, who were attempting to explain climate changes on Earth in terms of an increase in solar output due to the addition of accretion energy as the Sun passed through an interstellar cloud.  Shortly thereafter the same authors proposed 
hydrodynamic accretion from interstellar gas as a solution to the paradox of the extremely short lifetimes of massive stars compared to geologic timescales, which at the time was considered ``an exceedingly embarrassing consequence of the present theory of stellar evolution" \citep{hl39b}.

The Hoyle-Lyttleton accretion model is parameterized by an object of mass $M$ moving at a speed $V_\infty$ through 
a uniform medium of density $\rho_\infty$.  Using ballistic orbits (neglecting pressure effects), \cite{hl39a} defined an accretion radius given by
\begin{equation}
R_a=\frac{2GM}{V_\infty^2},
\end{equation}
such that gas approaching the star with an impact parameter less than $R_a$ would collide on an accretion line behind the star and would be left with insufficient kinetic energy to escape the gravitational potential of the star.
The mass accretion rate is expected to be comparable to the mass flux through a circle of radius $R_a$ far upstream:
\begin{equation}
\dot M_{HL} = \pi R_a^2 V_\infty \rho_\infty.
\label{eqn:mdot}
\end{equation}
A thorough review of Hoyle-Lyttleton accretion theory and subsequent numerical studies is given by \citet{e04}.
We provide only a brief synopsis here.
\defcitealias{ft88}{FT}

This basic model of hydrodynamic accretion was first investigated numerically by \citet{h71}, who used 2D axisymmetric steady-state solutions to quantify
the mass accretion rate.   For an upstream Mach number of 2.4 he found an accretion rate of $0.88\dot M_{HL}$. 
Subsequent time-dependent 2D axisymmetric simulations have, for the most part, shown that axisymmetric hydrodynamic accretion is stable \citep{shima85,pom00}.  \citet{ftm87} found steady axisymmetric accretion flows provided  that the inner boundary condition was totally absorbing.  \citet{koide91} found steady solutions for low Mach number flows, but they found the accretion flow to be unsteady if the upstream Mach number was greater than 2.  These latter results, however, are likely plagued by poor resolution near the accreting object.  The authors attributed the unsteady flow to gas flowing towards the accretor within the accretion column, but slightly missing the accretor.  One might expect the gas in the accretion
column to be completely absorbed if a higher grid resolution near the accretor fully resolved the flow.

This picture of steady, axisymmetric accretion is altered in the context of an x-ray binary system in which a compact object is accreting from the stellar wind of a companion star.  In this case gradients in both density and velocity of the flow past the compact star will break the axisymmetry, and in particular may lead to the accretion of angular momentum as well as mass \citep{is75,sl76}.
\citet{mis87} used 2D hydrodynamic simulations to study the dynamics of such wind accretion in a close binary system.  They presented some models
for which the flow near the accreting star sporadically changed its direction of rotation - a phenomenon the authors referred to as a `flip-flop flow.'

Shortly thereafter, numerical studies of the more idealized problem of Hoyle-Lyttleton accretion in 2D planar flow \citepalias{ft88} found 
a similar behavior,  with the accretion wake flipping from side to side.  The result of this flip-flopping is
the formation of temporary accretion disks spinning in alternating directions, with a burst of mass accretion when the direction of rotation is flipped.
This phenomenon has been seen in many subsequent simulations, and has come to be known as  the `flip-flop' instability.

While Hoyle-Lyttleton accretion is a very idealized problem, it is important to note that the flip-flop instability was first seen in more 
realistic simulations of gas flow within a binary star system \citep{mis87} and in subsequent simulations that incorporated several key physical
processes including x-ray heating and radiative cooling \citep{bkft90}.  Although in the latter case it was found that strong Compton heating due to a high
x-ray luminosity of the accreting star suppressed the flip-flop instability.  However, both of theses studies of accretion in binary systems were 
restricted to 2D.  The increase in density and pressure as the accretion flow approaches the compact star will be different in three dimensions
compared to two, raising the possibility that the stability of 3D accretion may differ from these 2D results.
The only published simulations of 3D Hoyle-Lyttleton accretion with a reasonably small accretor were a series of 
studies using a nested cartesian grid \citep{ra94,r99}.  The accretion in these 3D studies was found to be unstable, but the evolution
did not resemble the flip-flop behavior found in 2D planar flow.  In particular, the amplitude of variations remained considerably smaller than
that found in 2D, and no evidence of steady quasi-Keplerian disk flow was ever seen.  Nonetheless, some of the 3D models reported in
\citet{ra94} did exhibit a quasi-periodic swinging of the accretion shock leading to brief epochs of strong rotational flow near the accretor.
This behavior was particularly evident for the models with the smallest accretors.  
Although these 3D runs used nested grids to improve spatial resolution
near the accreting star, each nested grid was only $32^3$ zones.  This resolution 
is relatively low compared to that used in many of the 2D simulations exhibiting the flip-flop, and may have contributed 
to the lack of detection of the flip-flop instability in 3D.  

Independent of the question of whether adiabatic, 2D planar flow is an appropriate model for astrophysical systems,
the flip-flop instability itself has drawn significant scrutiny and criticism over the past two decades.
Many examples of 2D planar accretion simulations have been presented in the literature, with mixed results as tabulated by \citet{fgr05}.
Several authors have pointed out various shortcomings of existing 2D planar Hoyle-Lyttleton simulations. 
These include, but are not necessarily limited to, poor numerical resolution \citep{zwn95}, inaccurate algorithms \citep{pom00}, 
inappropriate boundary conditions upstream and at the surface of the accretor \citep{zwn95,blt97}, 
lack of conservation of angular momentum \citep{shima98}, and an unrealistically large accretor. 
These criticisms continue to cast doubt and confusion over the existence of the flip-flop instability.

In this paper we present an in-depth study intended to provide a 
better understanding of the properties of the flip-flop instability.
We take care to address the many issues 
raised about previous numerical simulations and we use the increased 
availability of computing power to more fully explore parameter space.  
We describe our numerical method in Section 2, building from the original flip-flop simulations of \citetalias{ft88}.
We present our results in Section 3,  where we show the flip-flop instability is a true overstability of the accretion wake, and provide a summary of our conclusions in Section 4.  

\section{Computational Method}

We use the time-dependent hydrodynamics code VH-1 to solve the Euler equations for an ideal gas characterized by an 
adiabatic index $\gamma$ in 2D polar coordinates with a point source of gravity at the coordinate origin.
This code is similar to the code used by \citetalias{ft88}. Both codes are based on the Piece-wise Parabolic Method \citep{cw84}, but the code used here uses the Lagrange-remap formulation while the code used by \citetalias{ft88} used the direct-Eulerian formulation.  None-the-less, we expect very similar results from these two codes.

\begin{figure}[!htp]
\begin{center}
\includegraphics[width=3.5in]{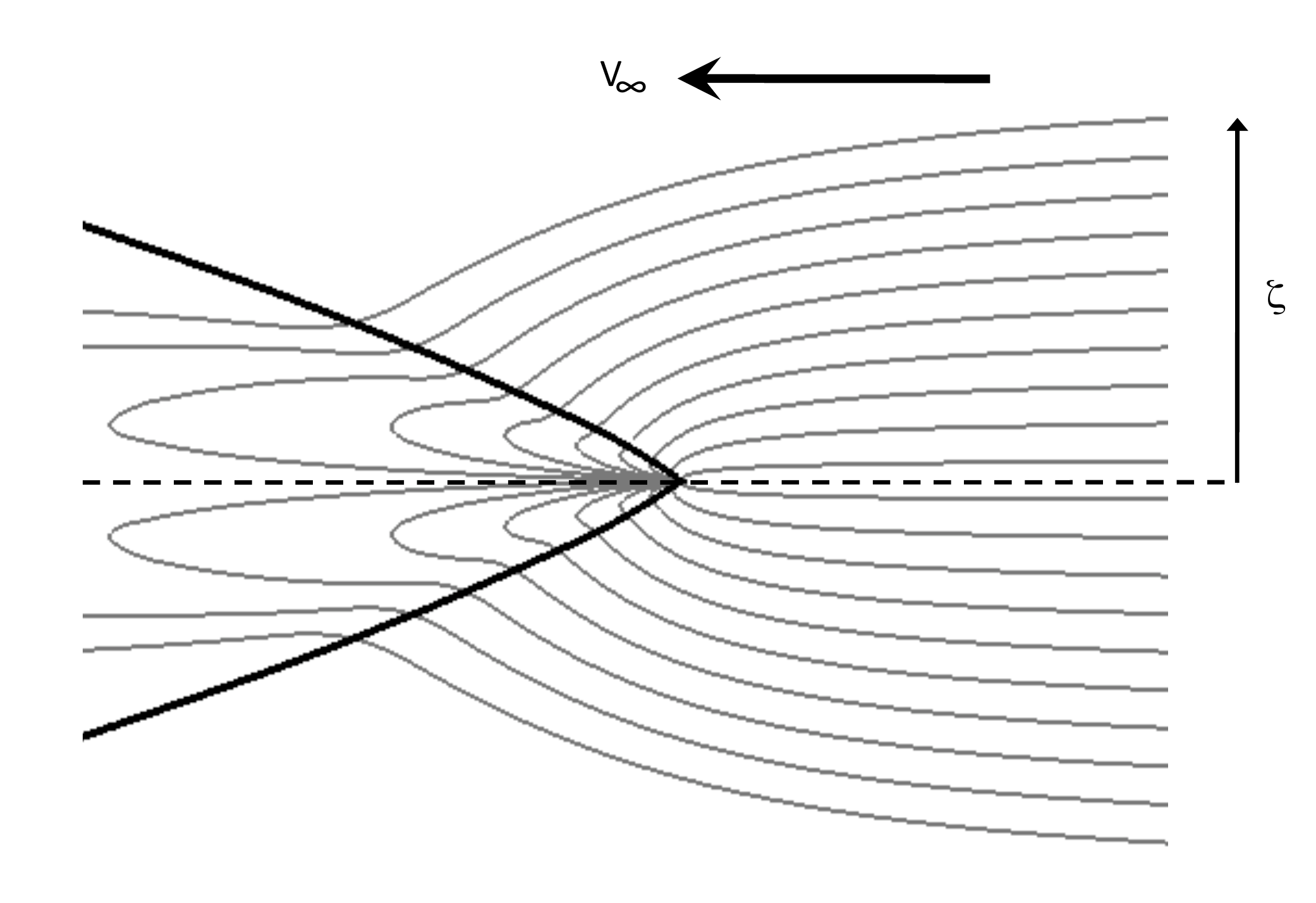}
\caption{The Holyle-Lyttelton accretion model with uniform flow at Mach 4 entering from the right 
and passing a point source of gravity, which is located at the apex of the shock cone marked 
by the heavy solid line.  The gray lines are streamlines as computed in the steady-state initial conditions
described in Section 2.3.  The outermost streamlines
correspond to an upstream impact parameter of $\zeta$ slightly greater than $R_a$. 
The parameters of the model shown here are $\gamma = 4/3$ and  $R_s = 0.0125\, R_a$}
\label{fig:schematic}
\end{center}\end{figure}

To this end, we begin by presenting results for a simulation set up to mimic a simulation in \citetalias{ft88}, 
using the same grid, boundary conditions and simulation parameters. Specifically, we use a numerical grid with 100 radial zones and 200 azimuthal zones. The inner radius corresponding to the surface of the accreting star is $R_s = 0.037\, R_a$, the first zone has a width of $\Delta R = 0.25\, R_s$, and the subsequent radial zones increase in width by a factor of 1.03. This puts the outer radius of the 
computational grid at $6.1\, R_a$. The flow at the upstream boundary is given a uniform velocity with a Mach number of 4. 
The gradients for all flow variables at the outer boundary
were set to zero for the downstream half of the grid, with the further condition that the radial velocity remained positive.
We imposed an absorbing boundary condition at the surface of the accreting star by setting
the density and pressure to very small values inside of this boundary \citep{h71}.   
We note that the flip-flop instability has been shown to be relatively insensitive to the specific boundary 
conditions imposed on the accreting surface \citep{tf89,zwn95}.
We imposed a transverse density gradient at the upstream boundary of the form
\begin{equation}
\rho=\rho_\infty\, (1+\frac{r\sin\phi}{R_a}\epsilon),
\label{eqn:epsilon}
\end{equation}
with $\epsilon = 0.005$.

\begin{figure}[!htp]
\begin{center}
\includegraphics[width=3.5in]{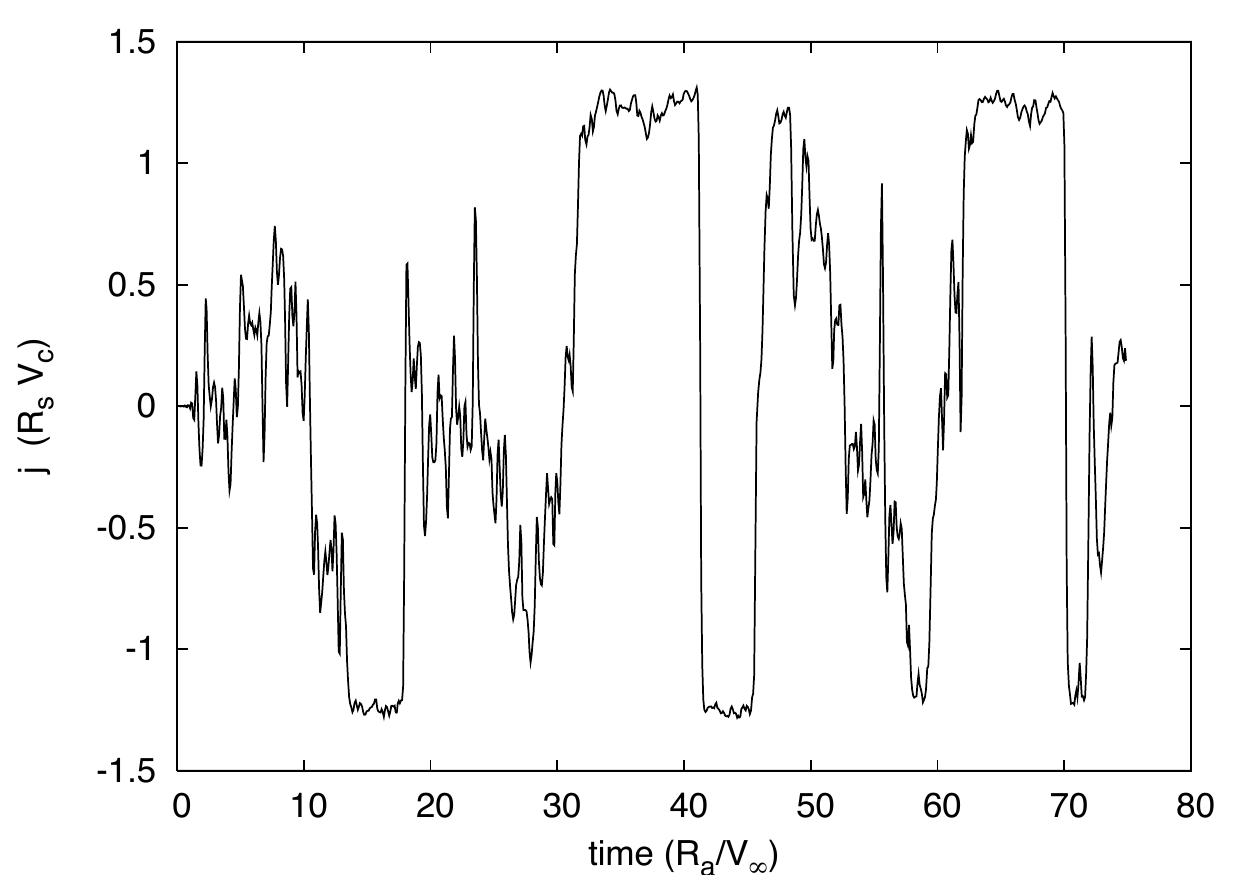}
\caption{The time evolution of the specific angular momentum of the accreted gas in our initial simulation with the standard VH-1 
hydrodynamics code is qualitatively similar to the
results of FT.  The flip-flop instability is characterized by episodes of disk accretion with alternating directions of rotation.}
\label{fig:bruce}
\end{center}\end{figure}

The results of this simulation, shown in Figure \ref{fig:bruce} as a trace of the specific angular momentum of the accreted gas, 
is qualitatively similar to the corresponding simulation presented in \citetalias{ft88}.  (Note that we normalize the
specific angular momentum to that of a circular orbit at the radius of the accretor, $R_s V_c = \sqrt{GMR_s}$,
so that the saturation value when a temporary accretion disk has
formed is independent of the accretor radius.)  In particular,
the accretion is marked by phases of disk accretion with relatively constant specific angular momentum corresponding roughly to that
of a circular orbit at the surface of the accretor.  These disk phases occur for both directions of rotation, with sometimes very abrupt
transitions between the two.

Given the dramatic increase in available computing power in the two decades since the discovery of the flip-flop instability, we are 
at liberty to run a large number of simulations to refine our computational model, to run at higher spatial resolution,
and to explore the parameter space more completely.   In the following subsections we delineate the changes made 
to this initial numerical model in order to arrive at what we consider to be high-fidelity hydrodynamic simulations demonstrating 
the exponential growth of the flip-flop instability.

\subsection{Hydrodynamic Algorithm}

A serious shortcoming of our initial model was the presence of density inhomogeneities in the upstream flow due to numerical noise.  
The source of these numerical fluctuations was traced to the conservation of total energy at the expense of 
ensuring constant entropy in the upstream supersonic flow traveling at various angles with respect to the computational grid.  
To reduce errors in internal energy created by subtracting kinetic energy from the conserved total energy,
we modified the algorithm to remap internal energy rather than total energy when the Mach number of the flow exceeds 3.0.  
This modification still provides for accurate shock jump conditions because the original conservation of total energy 
is used in zones containing a shock, but it now allows one to model uniform supersonic (Mach = 4) 
planar flow across a cylindrical polar grid with minimal numerical artifacts.

\subsection{Upstream Boundary Conditions}

Assuming a uniform planar flow at the upstream boundary neglects any gravitational effects of the flow from infinity to that boundary.  As a result, the flow near the accretor becomes dependent on the chosen radius of the outer boundary \citep{blt97}.  A more realistic boundary condition is to assume a ballistic trajectory from infinity to the outer boundary \citep{mss91,blt97,shima98}.  
The resulting flow velocity at a given point ($r,\phi$) in space is then \citep{bk79}
\begin{equation}
V_r= -V_\infty \sqrt{1 + \frac{R_a}{r} - \frac{\zeta^2 }{r^2}}
\end{equation}
\begin{equation}
V_\phi= V_\infty \frac{\zeta}{r} 
\end{equation}
where $\zeta$ is the impact parameter at infinity of the stream line passing through that point, 
\begin{equation}
\zeta=\frac{1}{2} \left[ r\sin\phi + \sqrt{r^2\sin^2\phi - 2rR_a(1+\cos\phi)}\right].
\label{eqn:zeta}
\end{equation}

One can then find the local gas density using mass conservation, equating the mass flux through a region bounded by $\zeta$ and $\zeta + d\zeta$ at infinity to the mass flux through a region bounded by $r$ and $r + dr$ for a point along the same stream line.  For the 2D planar geometry used in this paper the result is
\begin{equation}
\rho=\rho_\infty \, \frac{\zeta}{2\zeta-r\sin\phi}.
\end{equation}
Once the density is known, the pressure can be found by assuming the flow is isentropic.  
The flow variables upstream of the accretion shock in our simulations remain within 0.5 percent of this 
analytic solution for an upstream Mach number of 4.

\subsection{Steady Initial Conditions}

Although often not explicitly stated, previous simulations were typically 
started from an initial state of perfectly uniform planar flow.  The early 
evolution is then dominated by the formation of a tail shock along the symmetry axis behind the accretor and the subsequent 
expansion of the region of shocked gas.
This highly variable initial flow masks the origin of any possible instability, and indeed makes it impossible to assess whether the 
observed behavior is a true instability in the sense that an initial equilibrium state will tend to evolve away from equilibrium.

We address this problem by creating a steady-state solution for initial conditions.  We evolve our model on only half of the grid, 
enforcing reflection symmetry
about the line of centers.  These `half' simulations are initialized with the ballistic flow model and allowed to evolve for several flow
crossing times, during which the accretion shock settles into a steady state.  These relaxed solutions, an example of which
is shown in Figure \ref{fig:schematic}, are then used to initialize the
full 2D simulations.

\subsection{Downstream Boundary Conditions}

\begin{figure}[!htp]
\begin{center}
\includegraphics[width=3.5in]{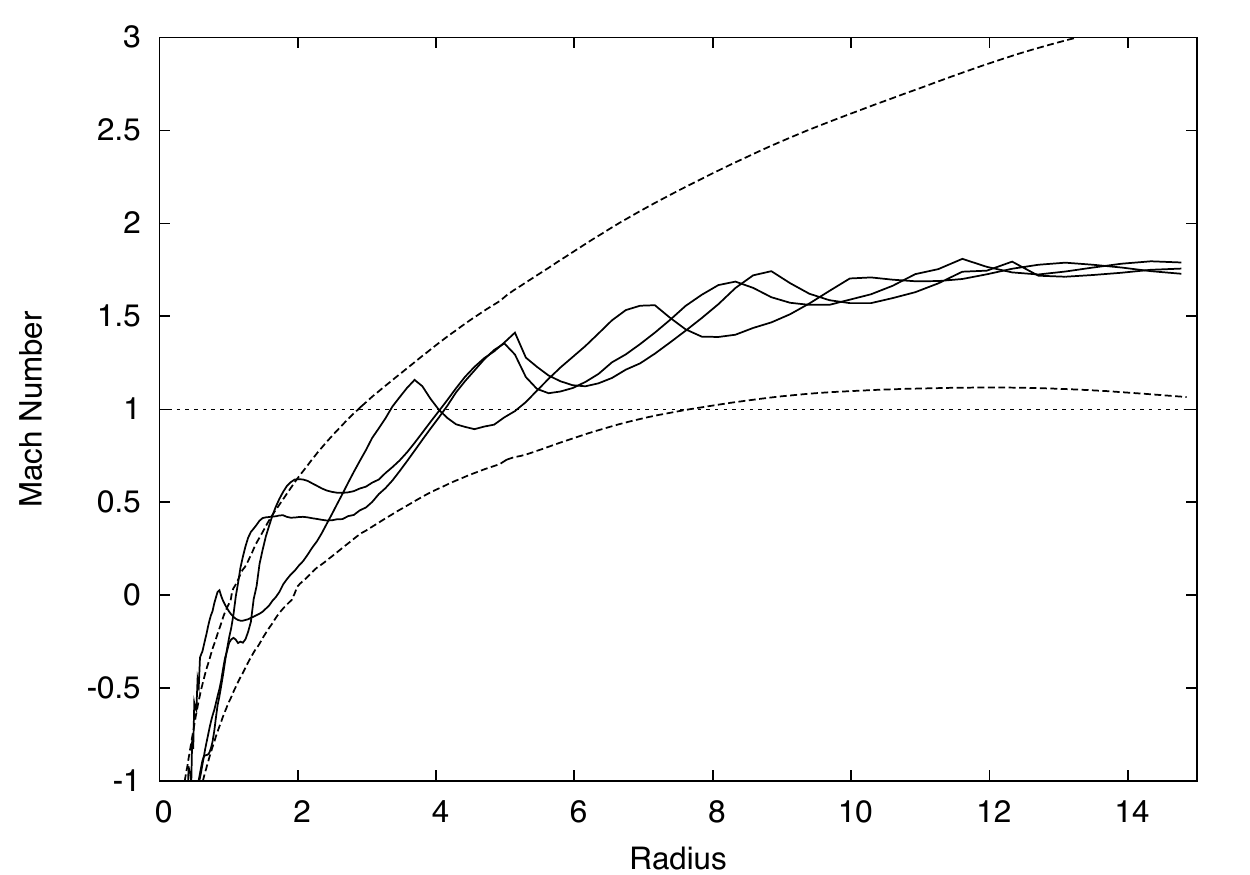}
\caption{Mach number of the outward radial flow inside the accretion wake as a function of distance from the accretor (in units of $R_a$).  
The two dashed lines
are from a steady-state solution, tracing the Mach number along angles of 7 and 14 degrees from the axis.  The solid lines show the minimum
Mach number inside the wake at different times in a simulation of unstable accretion flow.}
\label{fig:downstream}
\end{center}\end{figure}

The downstream boundary condition in previous numerical simulations has traditionally been a condition of zero gradients, allowing flow
to smoothly move off the grid.  This is a reasonable approach if the flow is moving supersonically off the grid, but can lead to spurious
reflections if the flow across the boundary is subsonic.  To investigate the downstream conditions in the accretion wake we examined the Mach
number of the flow both in our steady-state initial conditions and in the time-dependent unstable flow.  In the steady-state flow the radial velocity 
on the symmetry axis does not become positive until $\sim [width=3.5in]\,R_a$,
so we focus our attention to flow slightly off this axis.  In Figure \ref{fig:downstream} we plot
the Mach number of the radial flow as a function of distance from the accretor for angles of 7 and 14 degrees from the symmetry axis.  For 
reference, the shock is at an angle of about 21 degrees far downstream.  At 7 degrees from the axis, the flow exceeds Mach 1 only after 
traveling $\sim 8 R_a$, and even then it barely rises above unity.  We also show the minimum Mach number inside of the accretion wake
for a simulation subject to the flip-flop instability.  The unsteady flow falls in an envelope defined by the 7 and 14 degree traces.
Setting the outer radius to greater than 10 $R_a$, as done by \citet{blt97}, ensures that the gas will
propagate supersonically off of the grid under most circumstances, thereby minimizing wave reflection from the boundary.

\subsection{Conserve Angular Momentum}

\citet{shima98} demonstrate that the dynamics of the temporary accretion disks formed during the flip-flop instability are less
dependent on the underlying computational mesh if the hydrodynamic algorithm strictly conserves angular momentum rather
than linear momentum.  This result is not unexpected given that during the disk epochs 
mass accretion is driven by spiral shocks within a roughly Keplerian flow, and any mass accretion is closely tied to 
the transport of angular momentum.  To this end we made a minor change to the PPMLR algorithm 
used by VH-1 in order to explicitly conserve angular momentum.  This code modification has been used to quantify disk accretion
driven by spiral shocks  \citep{bl00}, measuring effective values of the disk $\alpha$ parameter as small as $10^{-3}$.
 
\subsection{Spatial Resolution}

The computational grid used by \citetalias{ft88} resulted in zones with extreme aspect ratios (8:1) near the surface of the accretor, while
other researchers used a single zone to represent the accretor \citep{mss91}.  More recent work has attempted to improve this
shortcoming by running identical simulations with increasing numerical resolution \citep{blt97} or by applying adaptive mesh
refinement to more accurately resolve the flow near the accretor \citep{zwn95}.
We chose to use a grid similar to that of  \citetalias{ft88},
namely a uniform grid in $\phi$ and a logarithmically increasing grid in $r$.  However, we followed the approach of \citet{blt97} and chose
parameters that
kept the aspect ratio of all zones roughly square.  This means both that $\Delta r/r$ is constant throughout the grid, and also that $\Delta r$ is much smaller than the radius of the accretor given that we use a minimum of 200 zones in the angular direction and we require that $\Delta r$ be comparable to $r\Delta\phi$.

\section{Results}

\begin{figure}%[!htp]
\begin{center}
\includegraphics{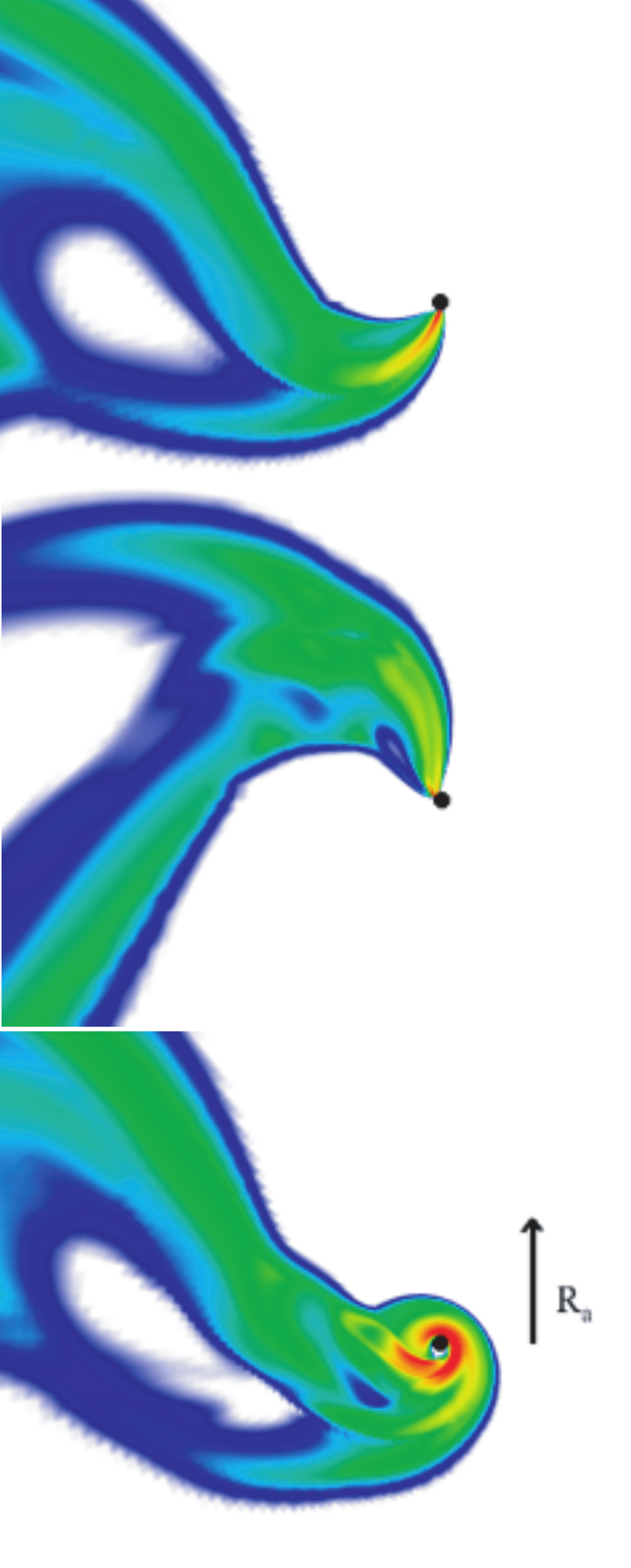}
\caption{The flip-flop instability is shown in this time series of the gas density from our standard model.  
The early evolution seen in the first two frames is characterized by the swinging of
the accretion shock from side to side, with ever increasing amplitude.  Eventually the amplitude grows to the point where the accretion shock
flips over the leading edge of the accretor, as seen in the last frame.  During these phases when the accretion shock has wrapped around
the accretor, the gas interior to the accretion shock forms a temporary accretion disk with nearly Keplerian flow.}
\label{fig:Aso08_seq}
\end{center}\end{figure}

For our standard model (A in Table 1) we chose parameters comparable to that of  \citetalias{ft88}, namely an upstream 
Mach number of four, $R_s = 0.037\,R_a$, and $\gamma = 4/3$.  We
did not impose any asymmetry in the upstream flow.
This simulation does exhibit the flip-flop instability (Figure ~\ref{fig:Aso08_seq}), however the oscillation of the 
accretion wake is less abrupt than found in the original simulations of  \citetalias{ft88}.  Much of the evolution is characterized by 
the accretion wake swinging back and forth, but remaining attached to the accretor.  Only for brief epochs does the shock wrap
around the leading edge of the accretor and `detach', resulting in a temporary accretion disk as seen in the last frame of 
Figure \ref{fig:Aso08_seq}.  Moreover, the flow never exhibits an abrupt transition from a disk of one rotation directly to a
counter-rotating disk.  Instead, the disk disappears and the accretion wake swings back and forth one or more times before
forming another disk.

\begin{figure}[!hbp]
\begin{center}
\includegraphics[width=3.5in]{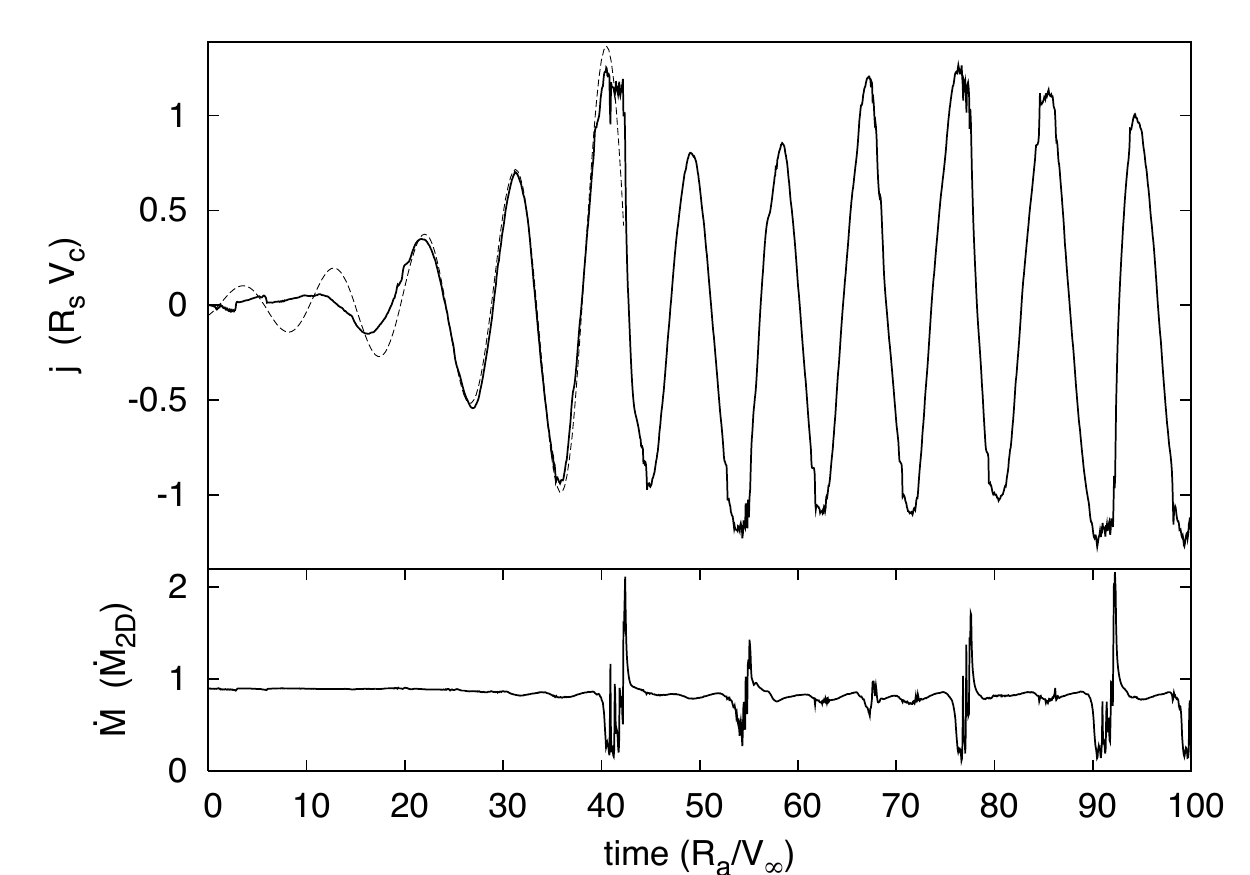}
\caption{The growth of the flip-flop instability from an initially equilibrium state is illustrated by the evolution of the specific angular momentum,
$j$,  of the accreted gas.  The dashed curve in the top panel is the analytic fit described in equation (9).
The mass accretion rate, shown in the bottom panel, is relatively constant except during brief periods of time when a disk forms and the mass
accretion rate drops substantially.  This low accretion phase is always followed by a large jump, when the mass in such a disk is 'flushed' onto the accretor resulting in a momentary spike in mass accretion.}
\label{fig:Aso01chart}
\end{center}\end{figure}

The mass accretion rate and the specific angular momentum of the accreted gas are shown in Figure \ref{fig:Aso01chart}.
The mass accretion rate is normalized to the 2D planar equivalent of equation (\ref{eqn:mdot}):
\begin{equation}
\dot M_{2D} = 2\, R_a \rho_\infty V_\infty.
\end{equation}
During the initial evolution, before the oscillations of the accretion shock grow to the point of driving the shock around the leading edge 
of the accretor, the mass accretion rate remains at a roughly constant value of $0.9 \,\dot M_{2D}$.  This is in good agreement with the 
original calculation of \citet{h71} for axisymmetric Hoyle-Lyttleton accretion as well as subsequent simulations of planar accretion 
flow \citep{koide91}.  During the brief epochs of disk accretion, the mass accretion rate is both
substantially lower and highly variable.  The termination of each disk phase is marked by a sudden spike in the mass accretion rate as the
bulk of the mass that had been orbiting in a quasi-Keplerian accretion disk is rapidly accreted on the relatively short free-fall timescale.

To quantify the growth of the instability we fit the early evolution (before the oscillations reach a maximum amplitude) 
with an exponentially growing sinusoid:
\begin{equation}
\dot j(t) = j_0 \, e^{\omega_r t} \cos (\omega_i t)  
\label{eqn:jdot}
\end{equation}
The fitted function shown in Figure \ref{fig:Aso01chart} corresponds to an oscillation frequency of $\omega_i = 0.68$ and an exponential
growth rate of $\omega_r = 0.07$.  This corresponds to the amplitude approximately doubling each period of oscillation.  
Note that the oscillation period ($2\pi/\omega_i = 9.2$) is comparable to the Keplerian period at the accretion radius ($2 \pi \sqrt{2} = 8.9$). 
Eventually some of the peaks in the trace of $j(t)$ become choppy, indicating those points when an accretion disk has 
formed (Figure \ref{fig:Aso01chart}).

\begin{figure}[!htp]
\begin{center}
\includegraphics[width=3.5in]{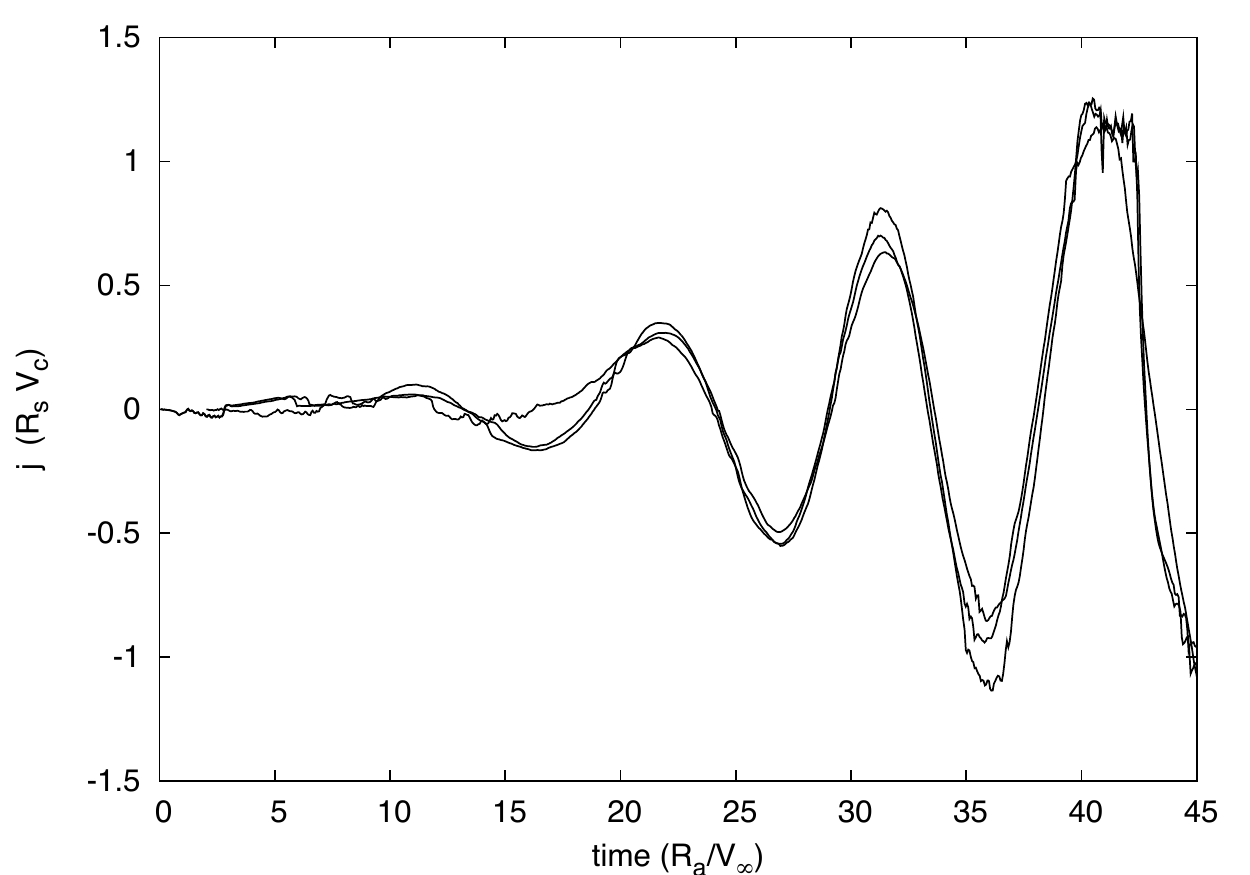}
\caption{The evolution of $j$ is shown for models A, B (Mach = 10), and C ($\epsilon=0.06$).}
\label{fig:compare}
\end{center}\end{figure}

These basic results were found to be relatively insensitive to the Mach number of the flow or the presence of a small asymmetry in 
the upstream flow.  
The original discovery of the flip-flop instability came from simulations in which a density gradient was 
imposed on the ambient medium.  To see if the instability is affected by small gradients, we ran a simulation (model C) with
a transverse density gradient as in equation (\ref{eqn:epsilon}), but with $(r\, \sin\phi)$ replaced by $\zeta$ as given in equation (\ref{eqn:zeta}).
We used an asymmetry value of $\epsilon = 0.06$.  We also ran a simulation (model B) with no
imposed asymmetry but with a higher Mach number of 10.  The results for these runs are shown in Figure \ref{fig:compare} in the form of the evolution
of the accreted specific angular momentum.  In each case the growth of the flip-flop instability was similar to the canonical run with $\epsilon=0$ and a Mach number of 4.  The growth rate of the instability in the higher Mach number flow was roughly 25\% higher, but otherwise the three simulations were very similar.

\begin{figure}[!htp]
\begin{center}
\includegraphics[width=3.5in]{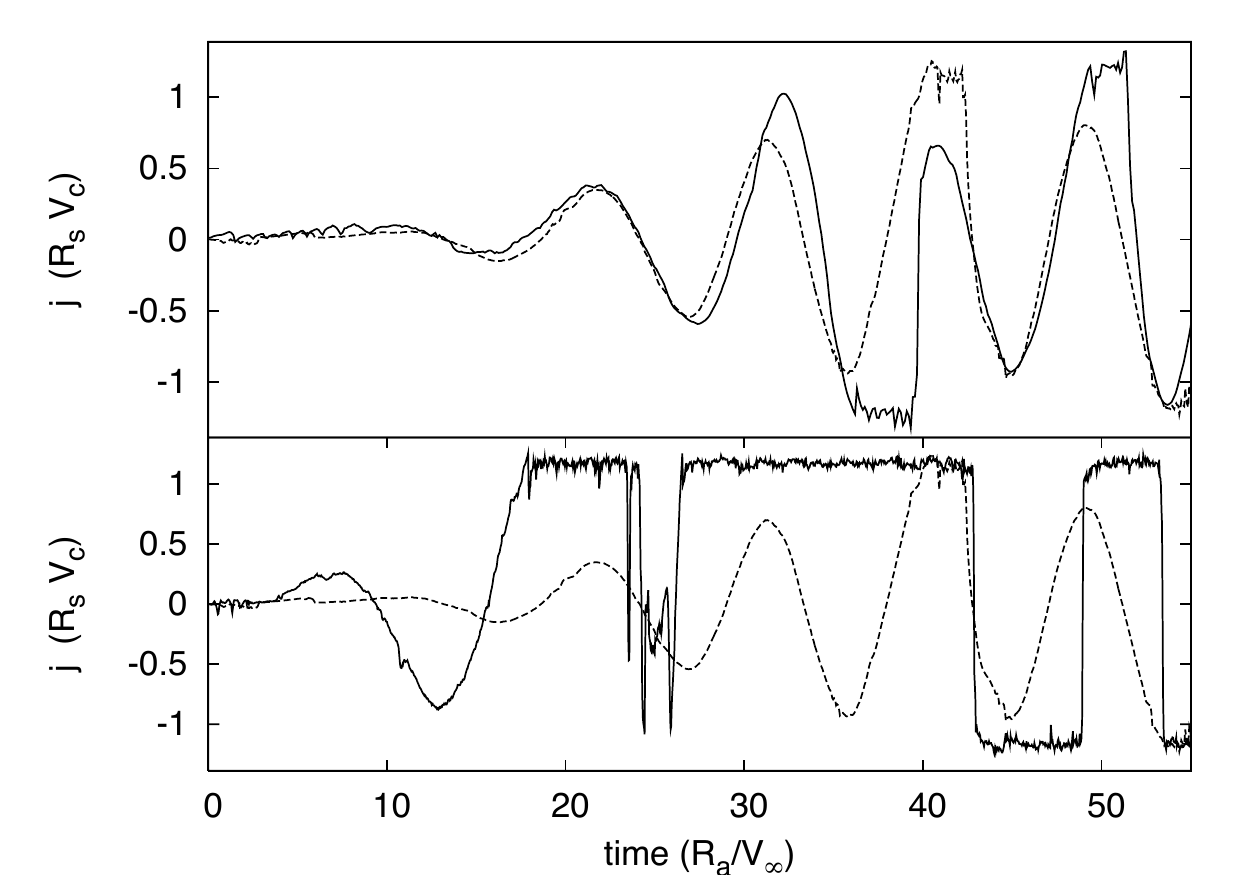}
\caption{The evolution of $j$ is shown for an accretor of radius $R_s = 0.0125\,R_a$ in the top plot, along with $j$ for the 
canonical model with $R_s = 0.037\,R_a$ (dashed line).  The bottom plot shows $j$ for an accretor of radius $R_s = 0.0037\,R_a$,
again compared against the canonical model. }
\label{fig:radiij}
\end{center}\end{figure}

For applications to accreting neutron stars and black holes, we are interested in values of $R_s$ that are far smaller
than our canonical value of $0.037\,R_a$.  Most simulations in the literature are limited to relatively large accretors
because decreasing $R_s$ becomes computationally very expensive.  
The timestep in these simulations is set by the Courant condition near
the surface of the accretor, and hence scales as $R_s^{3/2}$.  Some of our extended runs with small accretors ran
for more than $10^7$ time steps.  In addition to our canonical model 
we present results for models with an accretor radius of $0.0125\,R_a$ and $0.0037\,R_a$ in Figure \ref{fig:radiij}.  

A smaller accretor is more unstable.  The growth rate is larger and the flow forms an accretion disk quicker.  In the intermediate case the 
growth rate was roughly 50\% higher and the first disk was formed half an oscillation earlier.  In the smallest accretor case the growth rate
was more than twice as high as the canonical run and the first disk formed in half the time - after only one oscillation of the accretion shock.
The accretion disk
also persists longer for smaller accretors.  In the canonical model a temporary disk is formed for only brief epochs (typically
one flow time across the accretion diameter) with little impact on the quasi-periodic behavior of $j(t)$.  
In the case of the smallest accretor a given disk phase
can last for several flow times, with very abrupt transitions between counter-rotating disks.

\begin{figure}[!htp]
\begin{center}
\includegraphics{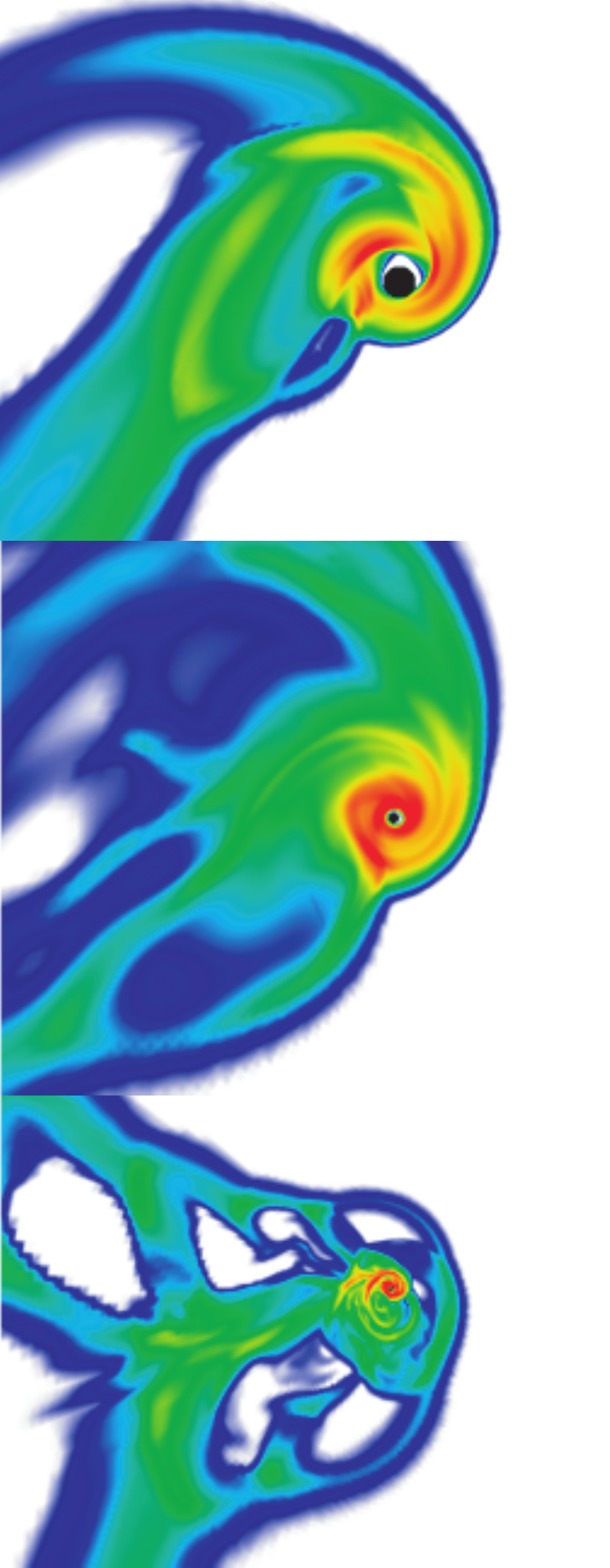}
\caption{Disk structure for three different sizes of the accretor, models A, D, and E.  The vertical size of each image is $R_a$.  
The density scale is different in each of the three images to allow similar detail to be seen in each.}
\label{fig:disks}
\end{center}\end{figure}

The structure of the temporary accretion disks as a function of accretor size is illustrated in Figure \ref{fig:disks}.
One can see a general trend of decoupling between the spiral shocks of the accretion disk and the leading bowshock as the 
size of the accretor is reduced.  In model A with the largest accretor the leading bowshock is a direct extension of a spiral shock
reaching down to the surface of the accretor.  In the intermediate model the spiral shock wraps at least one full revolution around
the accretor before it becomes a part of the external bowshock.  In the case of the smallest accretor (model E), the spiral shocks can be traced out
to the bowshock, but it involves several windings.  The result in this last case is a disk-like accretion flow that is only weakly coupled
to the bowshock.

The oscillation period of the flip-flop instability in the small amplitude regime (before the formation
of any accretion disk) is relatively insensitive to the accretor radius, with values comparable to the Keplerian period at the accretion radius.  
The dependence of $\omega_i$ on $R_s$ is hard to quantify because $\omega_i$ decreases somewhat as the
amplitude of oscillation increases.  As a result, the value of $\omega_i$ derived from the simulation data depends on the time window within
which one fits the analytic function (eq. [\ref{eqn:jdot}]).  This window is forced to be early in the evolution for the small accretor case because of the rapid
growth, but a better fit is found if a later window is used in the runs with a larger accretor.  One can say, however, that the differences in $\omega_i$
for different values of $R_s$ are smaller than the variations in $\omega_i$ with oscillation amplitude.  

\begin{deluxetable}{clrrrr}
%\tabletypesize{\scriptsize}
\tablecaption{Simulation Parameters}
\tablewidth{0pt}
\tablehead{
\colhead{Model} & \colhead{$R_s/R_a$} & \colhead{Mach} & \colhead{$\gamma$} & \colhead{$\epsilon$} & \colhead{$\omega_r$}}
\startdata
A & 0.037 & 4   & 1.333 & 0 & 0.070    \\
B & 0.037 & 10 & 1.333 & 0  & 0.088     \\
C & 0.037 & 4   & 1.333 & 0.06 & 0.070     \\
D & 0.0125 & 4 & 1.333 & 0  & 0.104     \\
E & 0.0037 & 4 & 1.333 & 0  & 0.170     \\
F & 0.037 & 4   & 1.667  & 0  & 0.000     \\
G & 0.037 & 4   & 1.400   & 0   & 0.068    \\
H & 0.037 & 4   & 1.500   & 0  & 0.040    \\
I & 0.037 & 4   & 1.550  & 0   & 0.031     \\
J & 0.0125 & 4 & 1.600 & 0 & 0.011 \\
K & 0.00185 & 4 & 1.667 & 0 & 0.008 \\
\enddata
\end{deluxetable}

\begin{figure}[!htp]
\begin{center}
\includegraphics[width=3.4in]{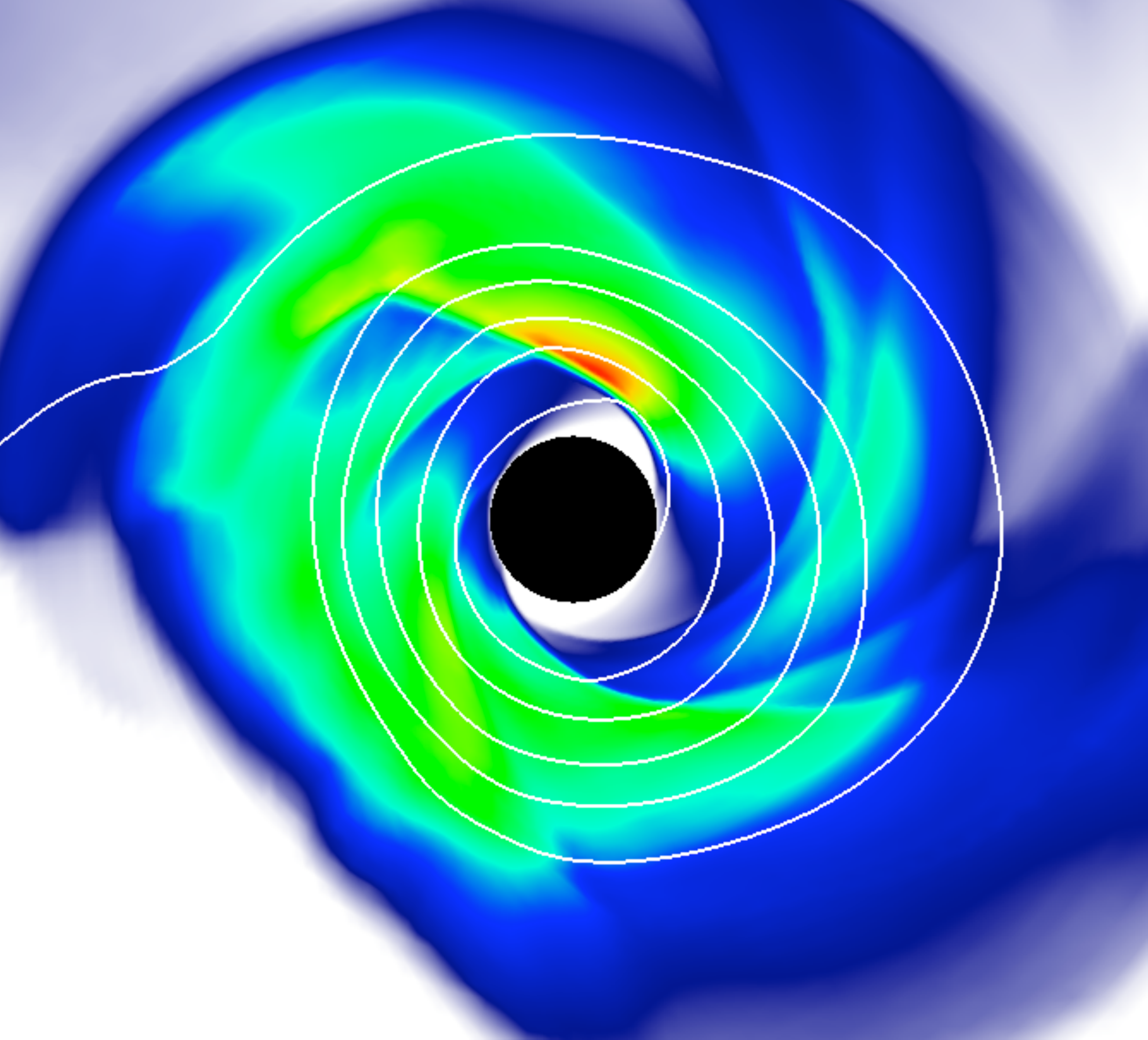}
\caption{A streamline from a snapshot of the flow in the simulation with $R_s = 0.0037\, R_a$ illustrates the elliptical nature of the orbits and
the influence of spiral shocks on the accretion of the orbiting gas.  The shading represents gas pressure.  The spiral shocks are
marked by discontinuities in pressure as well as slight kinks in the computed streamline.}
\label{fig:orbit}
\end{center}\end{figure}

The mass accretion during the disk phase is driven by spiral shocks \citep{bl00}.  This is illustrated in Figure \ref{fig:orbit}, where we 
show a streamline computed from the gas velocity at a fixed moment in time.  Note that this streamline is not the same as the path
taken by a fluid element over time, but it provides a better illustration of the effects of spiral shocks on the flow.  
When orbiting gas encounters a spiral shock, its orbital velocity is
decreased, resulting in the kinks seen in the streamline.
This process robs the gas of angular momentum, which is transferred to gas at larger radii through the pressure gradient behind the shock.
The loss of angular momentum in turn leads to an inward radial drift, and eventually accretion onto the central object.
The specific angular momentum of the gas accreted during a disk phase is  $j\sim 1.18$, independent of the size of 
the accretor.  This is slightly larger than the Keplerian value of unity at the surface of the accretor, consistent with the eccentricity
of the orbiting gas near the accretor.

Finally, we found that the adiabatic index has a dramatic influence on this instability. 
We ran several simulations varying Mach number and density asymmetry 
with $\gamma = 5/3$ and $R_s = 0.037\,R_a$, and in all cases the oscillation of the accretion shock remained bounded.  
Typically the shock cone begins oscillation very early on but  the oscillations never grow beyond a 
relatively small amplitude and no accretion disk is ever formed.  This dependence on $\gamma$ is consistent with the results
of \citet{zwn95}.  They ran a simulation with a homogeneous upstream flow and $\gamma = 5/3$ and reported steady
flow for an absorbing inner boundary.  They also ran a simulation similar to \citetalias{ft88} with $\gamma = 4/3$ and
$\epsilon = 0.005$ and found the flip-flop instability, regardless of the inner boundary condition.  The authors attributed
this difference to $\epsilon$, but given the results reported here this difference is more likely due to $\gamma$.

\begin{figure}[!htp]
\begin{center}
\includegraphics[width=3.4in]{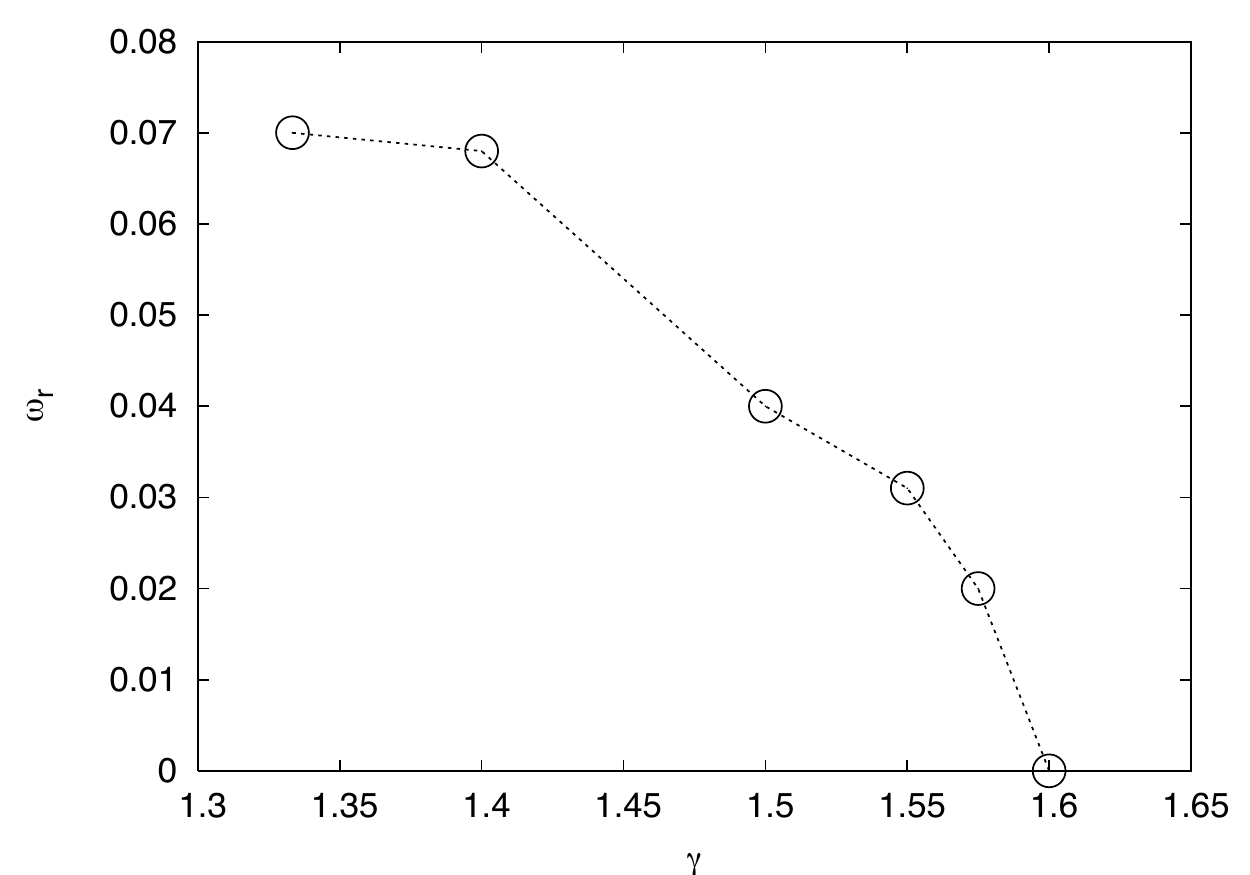}
\caption{The growth rate of the flip-flop instability is shown as a function of the adiabatic index of the gas, $\gamma$,
for $R_s = 0.037\, R_a$.  
The simulation with $\gamma=1.60$ exhibited small oscillations at a constant amplitude, but no evidence for growth.}
\label{fig:omega}
\end{center}\end{figure}

To further investigate this dependence on $\gamma$ we ran a series of simulations in which $\gamma$ was the only parameter 
changed.  To quantify the instability we fit $j(t)$ as in Figure \ref{fig:Aso01chart}.  The results, shown in Figure \ref{fig:omega}, show a monotonic
decrease in the growth rate with increasing $\gamma$.  By $\gamma=1.6$ the accretion wake is no longer unstable.  The wake still
swings back and forth, but the amplitude of those oscillations is only a few degrees and  it remains remarkably constant.

\begin{figure}[!htp]
\begin{center}
\includegraphics[width=3.4in]{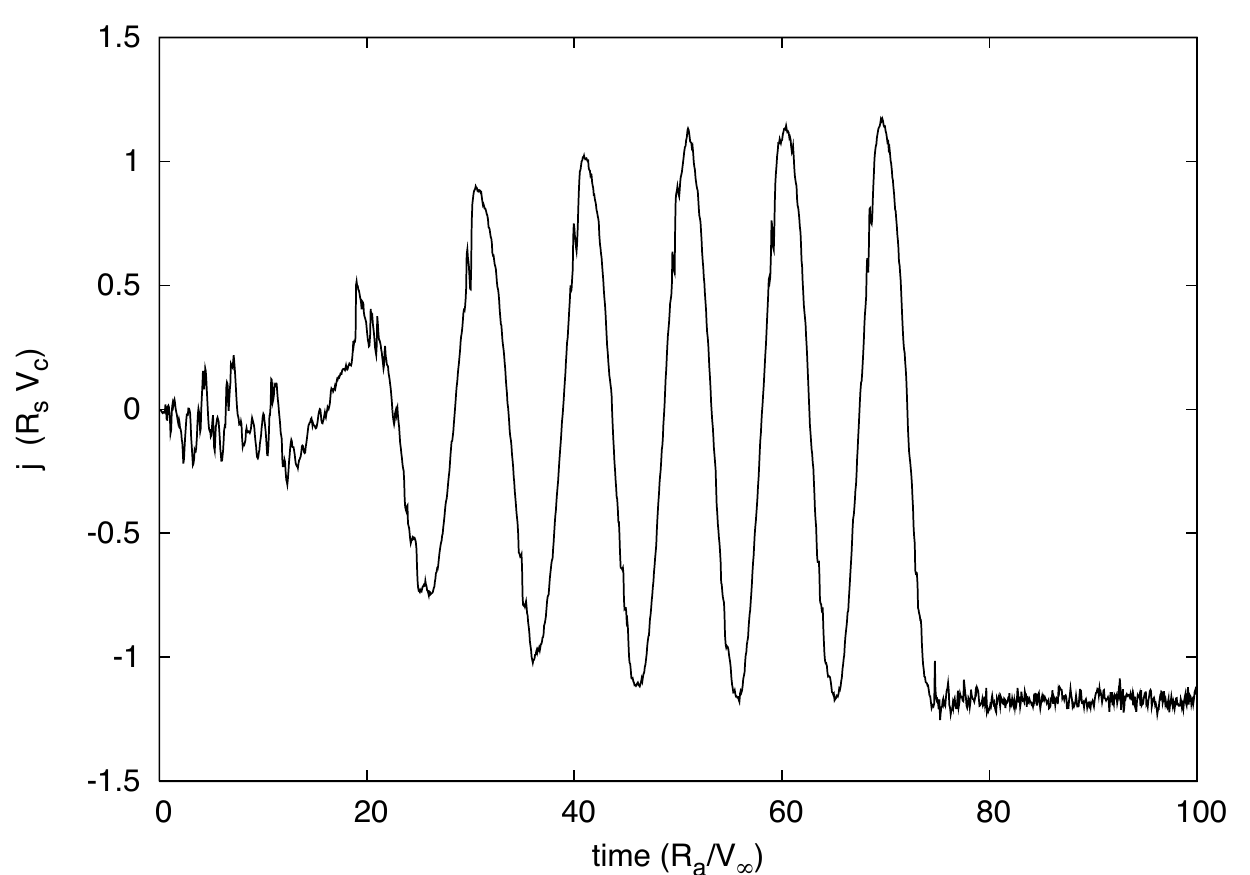}
\caption{The growth of the flip-flop instability for an accretor of radius $R_s = 0.00185\,R_a$ and
adiabatic index $\gamma=5/3$.}
\label{fig:flipped}
\end{center}\end{figure}

This critical value of $\gamma$, above which the flow is (at least marginally) stable, depends on the size of the accretor.  As shown above,
a smaller accretor produces a faster growth rate of the instability.  A smaller accretor also means a larger critical value of $\gamma$.  We 
ran simulations with $\gamma=1.6$ for smaller accretors (see Table 1) and found the flow to be unstable, with increasing growth rate for
decreasing $R_s$ as in the case with $\gamma=4/3$.  We also ran simulations with $\gamma=5/3$, searching to see if such flow is
unstable if the size of the accretor is sufficiently small.  With $R_s=0.0025\,R_a$, the amplitude of the the oscillating accretion shock
grew very slowly with $\omega_r = 0.006$, but then seemed to level off before the amplitude became big enough for the shock to flip over the leading
edge and produce a disk.  The growth rate was about the same for a slightly smaller accretor, $R_s=0.00185\,R_a$,  but as shown
in Figure \ref{fig:flipped} the amplitude was big enough to produce a flip after only six oscillations.

\section{Conclusions}

We have presented high-fidelity simulations of 2D planar accretion onto a compact object in order to better understand the 
flip-flop instability originally discovered by \citetalias{ft88}.  From these simulations we can draw the following conclusions:

1) First and foremost, we have shown that the flip-flop is a true
overstability, where a small displacement of an equilibrium accretion shock leads to oscillations with exponentially growing amplitude.

2) This instability does not require an asymmetry in the upstream flow; a result first suggested by \citet{mss91}.

3) The growth rate is only weakly dependent on the Mach number of the upstream flow, with higher Mach number producing slightly faster growth of the instability.

4) The growth rate increases with decreasing radius of the accretor.  

5) For a given accretor size, the stability of the flow is dependent on the adiabatic index of the gas, 
with the growth rate decreasing as $\gamma$ increases from 4/3.  Planar accretion is stable 
for $\gamma \ge1.6$ in the case of a large accretor ($R_s = 0.037\,R_a$),
while $\gamma=5/3$ is only unstable for the smallest accretor used in our simulations, $R_s = 0.00185\,R_a$.

\acknowledgments

This work was supported in part by NSF grant OCI-749248.  T. C. P. was supported in part by an
Undergraduate Research Award from North Carolina State University.  
The authors acknowledge the Texas Advanced Computing Center (TACC) at 
The University of Texas at Austin for providing computing resources that have contributed 
to the research results reported within this paper.  We thank the anonymous referee for pointing out
that the critical $\gamma$ for unstable flow might depend on the radius of the accretor.

\clearpage

\end{document}